\documentclass[sigconf]{acmart}
\usepackage{appendix,float}
\usepackage{array}
\usepackage{subcaption}
\usepackage{graphicx}
\usepackage{tikz}
\usepackage{wrapfig}
\usepackage[utf8]{inputenc}
\usetikzlibrary{calc, positioning, arrows.meta}
\usetikzlibrary{decorations.pathreplacing} 

\AtBeginDocument{%
  \providecommand\BibTeX{{%
    \normalfont B\kern-0.5em{\scshape i\kern-0.25em b}\kern-0.8em\TeX}}}

\copyrightyear{2025} 
\acmYear{2025} 
\setcopyright{acmlicensed}
\makeatletter
\renewcommand{\@authorfont}{\normalfont\large}
\makeatother
\begin{document}
\title{Generative AI for CAD Automation: Leveraging Large Language Models for 3D Modelling}
\author{Sumit Kumar}
\email{skumar18@student.gsu.edu}
\affiliation{%
  \institution{Georgia State University}
  \city{Atlanta}
  \state{GA}
  \country{USA}
}

\author{Sarthak Kapoor}
\email{sarkapoo@amazon.com}
\affiliation{%
  \institution{Amazon LLC}
  \city{Seattle}
  \state{WA}
  \country{USA}
}

\author{Harsh Vardhan}
\email{harsh.vardhan@vanderbilt.edu}
\affiliation{%
  \institution{Vanderbilt University}
  \city{Nashville}
  \state{TN}
  \country{USA}
}

\author{Yao Zhao}
\email{yaozhao16@temple.edu}
\affiliation{%
  \institution{Temple University}
  \city{Philadelphia}
  \state{PA}
  \country{USA}
}

\renewcommand{\shortauthors}{Kumar S., et al.}

\begin{abstract}
Large Language Models (LLMs) are revolutionizing industries by enhancing efficiency, scalability, and innovation. This paper investigates the potential of LLMs in automating Computer-Aided Design (CAD) workflows, by integrating FreeCAD with LLM as CAD design tool. Traditional CAD processes are often complex and require specialized sketching skills, posing challenges for rapid prototyping and generative design. We propose a framework where LLMs generate initial CAD scripts from natural language descriptions, which are then executed and refined iteratively based on error feedback. Through a series of experiments with increasing complexity, we assess the effectiveness of this approach. Our findings reveal that LLMs perform well for simple to moderately complex designs but struggle with highly constrained models, necessitating multiple refinements. The study highlights the need for improved memory retrieval, adaptive prompt engineering, and hybrid AI techniques to enhance script robustness. Future directions include integrating cloud-based execution and exploring advanced LLM capabilities to further streamline CAD automation. This work underscores the transformative potential of LLMs in design workflows while identifying critical areas for future development.
\end{abstract}


\maketitle
\section{Introduction}
The use of artificial intelligence (AI) has seen exponential growth across a wide range of domains, including finance, automotive \cite{kumari2024automatic}, healthcare \cite{kumar2023malaria}, engineering design \cite{vardhan2024sample}, Cybersecurity \cite{kumar2025cyber,neema2024rampart} and beyond, revolutionizing how tasks are automated, decisions are made, and innovations are realized. Among the most transformative advancements in AI is the rise of generative AI (Gen AI), which focuses on creating new content, designs, and solutions rather than merely analyzing existing data. Gen AI has opened up unprecedented possibilities, from generating realistic images and text to automating complex design workflows.

Within the realm of Gen AI, LLMs have rapidly emerged as a transformative impact across industries, reshaping how businesses operate, innovate, and compete. Organizations that have already integrated LLMs into their workflows are witnessing significant gains in efficiency, cost reduction, and scalability. Meanwhile, industries that have yet to adopt these models are actively investing in their development to remain competitive and future-proof their operations. As LLMs continue to evolve, their impact extends beyond automation, enabling new capabilities in decision-making, knowledge synthesis, and personalized user experiences. This paper explores how effectively LLMs are tackling design challenges, where their capabilities currently stand, and what the future might hold for their role in solving complex design problems.

CAD is the backbone of several industries such as manufacturing, construction, engineering, and design \cite{saxena2007computer}. We are heavily reliant on CAD for most of the development and prototyping. Sectors like mechanical engineering, aerospace, automotive, etc. use CAD to build physical objects \cite{ibrahim2003cad}. CAD tools like FreeCAD, AutoCAD, and SolidWorks \cite{matsson2023introduction} allow engineers to design highly precise 3D models, reducing time and costs associated with traditional prototyping. However, CAD modeling remains a complex, labor-intensive process requiring specialized expertise, particularly in scripting and parametric modeling. To operate the CAD tool, the operator should not only be proficient in design knowledge but also possess strong CAD scripting skills. CAD scripting requires a good understanding of several components, including familiarity with specific libraries such as the FreeCAD Part and Mesh modules, refining designs through trial and error, making incremental modifications, and debugging scripts. Even with strong expertise in both design principles and CAD scripting, the process remains time-consuming and iterative. Writing and debugging scripts to achieve precise parametric designs demands significant effort, often requiring multiple refinements to meet design specifications. This bottleneck not only slows down the design cycle but also limits accessibility for those without extensive programming experience.

With an increasing demand for rapid prototyping, generative design, and automated manufacturing, there is a pressing need for AI-driven intelligent approaches that can automate the entire CAD modeling workflow, from design generation to validation and error correction.

\section{Problem Formulation and Integration Architecture}
\label{sec:architecture}
For an end-to-end execution, we integrated LLM in loop with CAD tool and error-driven refinement and prompt evolution. Next, we will explain each section.
\subsection{Initial FreeCAD Script Generation Using LLMs}
LLMs such as GPT-4 operate as probabilistic generators of text. Given a natural language design description \( d \) and a structured prompt \( P_i \), which includes constraints, syntax rules, and guidelines for FreeCAD scripting, the LLM generates an initial Python script \( S \). Mathematically, we define the generation process as:

\begin{equation}
    S = \mathcal{G}(P_i, d)                
\end{equation}

where:

\renewcommand{\labelitemi}{--} 
\begin{itemize}
    \item \( S \) is the generated FreeCAD Python script.
    \item \( P_i \) represents the structured initial prompt that provides constraints, best practices, and FreeCAD-specific scripting requirements.
    \item \( d \) is the user-provided natural language design description, which defines the intended 3D model.
    \item \( \mathcal{G}(\cdot) \) represents the function that maps text inputs to Python scripts using the trained LLM.
\end{itemize}

Since LLMs operate by sampling from a probability distribution over possible sequences, the objective is to maximize the probability of generating an optimal FreeCAD script:

\begin{equation}
    \mathcal{G}(P_i, d) = \arg\max_{S} P(S \mid P_i, d, \theta)
\end{equation}

where:
\renewcommand{\labelitemi}{--} 
\begin{itemize}
    \item \( \theta \) represents the learned parameters of the LLM, encoding statistical relationships from its training data.
    \item \( P(S \mid P_i, d, \theta) \) is the probability distribution over all syntactically and semantically valid FreeCAD scripts.
\end{itemize}

This formulation implies that the generated script \( S \) is not deterministic but rather probabilistically sampled from the most likely completions under the model's learned weights.

\subsection{Execution and Error Handling}
Once the FreeCAD script \( S \) is generated, it is executed in headless mode within FreeCAD to generate the corresponding 3D model. However, due to the probabilistic nature of LLMs and potential inaccuracies in FreeCAD scripting logic, errors may occur. These errors include:
\renewcommand{\labelitemi}{--} 
\begin{itemize}
    \setlength\itemsep{0em}
    \item \textbf{Syntax Errors}: Invalid Python code due to incorrect indentation, missing imports, etc.
    \item \textbf{Geometric Inconsistencies}: Invalid Boolean operations, degenerate geometry.
    \item \textbf{Execution Failures}: Incorrect API calls, missing object dependencies.
\end{itemize}

The execution process can be mathematically represented as:

\begin{equation}
    E = \mathcal{F}(S)
\end{equation}

where:
\renewcommand{\labelitemi}{--} 
\begin{itemize}
    \item \( E \) is the error message returned by FreeCAD during execution.
    \item \( \mathcal{F}(\cdot) \) represents the FreeCAD execution function, which evaluates the script and detects errors.
\end{itemize}

If \( E = 0 \), the script runs successfully, and the 3D model is generated. Otherwise, iterative refinement is required.

\subsection{Iterative Refinement via Error-Driven Prompt Evolution}
LLM APIs are \textbf{stateless}, meaning they do not retain memory of previous interactions unless explicitly designed to do so (e.g., using embeddings, vector databases, or session-based memory) \cite{yu2023stateful}. Each API call is independent, and the LLM does not "remember" past conversations unless context is explicitly provided in the request. Therefore in the refinement process we are proving the initial user request in \( P_r \). When an error \( E \) is detected, the system must refine the generated script. However, directly modifying the script may introduce cascading issues (e.g., unintended changes, loss of user intent). Instead, we adopt a structured approach where the prompt itself is refined before regenerating the script.
The refined prompt \( P_r \) is constructed as:

\begin{equation}
    P_r^{(t+1)} = f(P_i, E^{(t)})
\end{equation}

where:
\renewcommand{\labelitemi}{--} 
\begin{itemize}
    \item \( P_r^{(t+1)} \) is the updated prompt after incorporating the latest error information.
    \item \( f(\cdot) \) is a refinement function that takes the original prompt \( P_i \) and the observed execution error \( E^{(t)} \) to generate a more constrained prompt.
\end{itemize}

The script is then regenerated as:

\begin{equation}
    S^{(t+1)} = \mathcal{G}(P_r^{(t+1)}, d, S^{(t)}, E^{(t)})
\end{equation}

where:
\renewcommand{\labelitemi}{--} 
\begin{itemize}
    \setlength\itemsep{0em}
    \item The LLM receives the refined prompt \( P_r^{(t+1)} \), along with the original description \( d \), the last generated script \( S^{(t)} \), and the observed error \( E^{(t)} \).
    \item The new script \( S^{(t+1)} \) is expected to minimally modify \( S^{(t)} \) to correct errors while preserving structural integrity.
\end{itemize}

This refinement process continues iteratively:
\begin{equation}
    S^{(t+1)} = S^{(t)} + \Delta S
\end{equation}

where \( \Delta S \) represents the minimal change required to eliminate the observed error.
\newline
\newline
The refinement process terminates when:
\begin{equation}
    E^{(t)} = 0 \quad \text{or} \quad t \geq T
\end{equation}

where:
\renewcommand{\labelitemi}{--} 
\begin{itemize}
    \setlength\itemsep{0em}
    \item \( T \) is a predefined maximum retry limit.
    \item If \( t \geq T \) and \( E^{(t)} \neq 0 \), the system fails gracefully by logging the error for manual intervention.
\end{itemize}

\subsection{Framework Design}
\begin{figure*}[ht]
    \centering
    \includegraphics[width=1\textwidth]{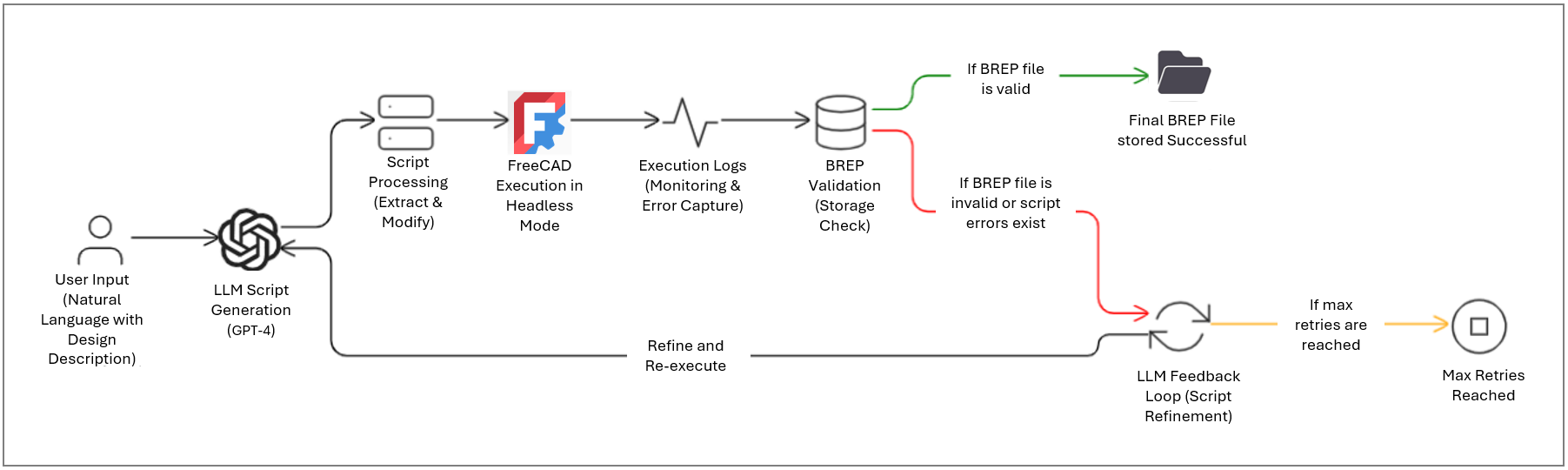}
    \caption{System architecture of the LLM-driven FreeCAD script generation framework. The workflow begins with a natural language input, which is processed by the LLM (GPT API + LangChain) to generate a FreeCAD script. The script is executed in FreeCAD's headless mode, and errors are fed back into the refinement loop until a valid 3D model is generated.}
    \label{fig:architecture}
\end{figure*}

The proposed framework for automating FreeCAD script generation using LLMs is illustrated in Figure~\ref{fig:architecture}. It describes the workflow, beginning with natural language input, progressing through its processing, and ultimately resulting in the generation of a 3D model. The framework leverages the GPT API for script generation and LangChain for prompt management \cite{auffarth2023generative} and iterative refinement, enabling the system to interpret design descriptions, generate scripts, and refine them based on execution feedback. The architecture consists of four key components described below.

This architecture emphasizes a closed-loop feedback mechanism, where errors drive the refinement process. This approach ensures that the final output aligns with the user's design intent while maintaining geometric and parametric validity. The iterative nature of the framework highlights the importance of adaptive prompt engineering and error handling to achieve reliable script generation. To balance computational efficiency, especially for complex problems, the system incorporates a maximum retry limit, ensuring that the process terminates gracefully after a predefined number of iterations. The use of BREP (Boundary Representation) to export 3D models ensures a precise geometric representation, as it can accurately capture complex shapes and features that STL (Stereolithography) often fails to represent \cite{gonzalez2024facilitating}. This makes BREP particularly suitable for CAD workflows and downstream applications such as 3D printing and manufacturing.

The framework consists of four key components, as illustrated in Figure~\ref{fig:architecture}:
\begin{itemize}
    \item \textbf{Natural Language Input}:  
    The process begins with the user providing a natural language description of the desired 3D model. This description includes details such as dimensions, geometric features, and constraints. The input serves as the foundation for the LLM to generate an initial FreeCAD script. This user input is further combined with the initial prompt to make sure we can get an efficient response from the LLM which can be directly used for modeling in the FreeCAD headless mode.

    \item \textbf{LLM with GPT-4 API and LangChain}:  
    The LLM, powered by the GPT-4 API, generates a FreeCAD script in Python based on the user's initial input and a predefined initial prompt. LangChain is used to manage the initial interaction with the GPT API, including prompt structuring and script generation. Specifically, LangChain's \texttt{PromptTemplate} ensures that the initial script includes all necessary imports, object validations, and FreeCAD-specific commands such as \texttt{Part.show()} and \texttt{FreeCAD.ActiveDocument.recompute()}. However, the refinement process, which involves updating the prompt based on error feedback from script execution, is handled manually by constructing a new prompt that includes the original user request, the terminal log, and the initial script. This refined prompt is then fed back into the GPT API for script correction. While LangChain is not directly involved in the refinement loop, its initial prompt structuring capabilities play a crucial role in ensuring the robustness of the generated scripts.

    \item \textbf{FreeCAD Execution Environment}:  
    Once the script is generated, it is executed in FreeCAD's headless mode, which allows for automated script evaluation without manual intervention \cite{wolf2020software}. During execution, the system checks for errors such as syntax issues, geometric inconsistencies, or invalid API calls. If errors are detected, they are logged and fed back into the refinement loop.

    \item \textbf{Iterative Refinement Loop}:  
    The refinement process is a critical component of the architecture. When errors are identified during script execution, LangChain constructs a refined prompt that includes the original user request, the terminal log (which combines both standard output and error messages), and the initial script. The terminal log provides a comprehensive record of the script’s execution, capturing both successful outputs (stdout) and error messages (stderr). While stderr is essential for identifying specific errors such as syntax issues, invalid API calls, or geometric inconsistencies, stdout offers valuable context about the script’s execution flow before the error occurred. This combination ensures that the LLM has sufficient information to diagnose and correct the issue effectively. The refined prompt (with terminal log) is then fed back into the GPT API to generate an updated version of the script. This process continues iteratively until the script executes successfully and produces a valid 3D model or until a predefined maximum number of retries is reached. By leveraging both stdout and stderr, the framework ensures a more robust and a full context-aware refinement process, ultimately improving the accuracy and reliability of the generated scripts.
    
\end{itemize}

The workflow is designed to minimize manual intervention while ensuring the accuracy and robustness of the generated scripts. By leveraging the GPT API’s ability to interpret natural language and LangChain’s capabilities in prompt management and refinement \cite{alto2024building}, the framework bridges the gap between high-level design descriptions and low-level CAD scripting.

\balance
\section{Experiment and Assessment }
\label{sec:casestudy}

To evaluate the effectiveness of LLM in generating FreeCAD scripts, we designed a structured set of 10 test cases with increasing levels of complexity. These experiments assess how well an LLM can generate functional CAD scripts that correctly execute within FreeCAD’s environment, ensuring syntactic accuracy, geometric validity, and parametric adaptability. The evaluation follows a complexity scale from 1 (simple shape creation) to 10 (advanced parametric modeling with constraints).

\subsection{Experimental Setup}
We used OpenAI’s GPT-4 as the primary LLM for generating FreeCAD scripts. Each test case was prompted using a structured format, explicitly defining shape dimensions, positioning, and constraints to minimize ambiguity. The generated scripts were then executed in FreeCAD’s headless mode, where they either successfully created the desired 3D model or failed due to errors. Failures triggered an iterative refinement process, where errors were analyzed and fed back into the LLM for script correction. 


\subsection{Prompt Complexity Scaling}
Each test case was structured to incrementally increase in difficulty, introducing additional modeling constraints and dependencies. Table~\ref{tab:prompt_levels} provides a summary of the ten test cases. At lower complexity levels (1-3), the LLM was tested on generating individual geometric primitives such as cubes, cylinders, and filleted cuboids. These prompts involved explicit shape dimensions and placements without interdependencies, making them straightforward for the model. At moderate complexity levels (4-7), prompts involved Boolean operations, parametric constraints, and hierarchical dependencies. The LLM was required to generate scripts that combined multiple objects through union and subtraction, ensuring the correct order of operations and shape dependencies. At higher complexity levels (8-10), test cases focused on intricate feature dependencies, requiring the LLM to generate involute gear profiles, parametric frames, and reinforcement ribs. These models demanded precise geometric constraints, ensuring modifications to one dimension correctly adjusted all related features. For the detailed prompt, refer to Appendix ~\ref{sec:exp_results}

\subsection{Performance Analysis}

To evaluate the performance of this automation framework, four primary metrics were recorded: (1)~first-attempt success or failure, (2)~the number of iterative refinements, (3)~total execution time (including re-prompting overhead), and (4)~the type of errors encountered.
Table~\ref{tab:results_summary} provides a concise overview of these findings.

\begin{table*}[ht]
\centering
\small 
\caption{Summary of Iterations, Execution Times, and Outcomes Across Ten Test Cases}
\label{tab:results_summary}
\resizebox{\textwidth}{!}{%
\begin{tabular}{l l c c l l}
\toprule
\textbf{Test} & \textbf{Shape/Task} & \textbf{Iterations} & \textbf{Time (s)} 
             & \textbf{Error Type} & \textbf{Outcome} \\
\midrule
1  & Basic cube (50$\times$50$\times$50 mm)       & 1           & 19.06  
   & None                      & Success, first attempt \\
2  & Cylinder (r=30 mm, h=80 mm)                  & 1           & 20.29  
   & None                      & Success, first attempt \\
3  & Filleted cuboid (100$\times$50$\times$30 mm) & 2           & 42.00  
   & Minor syntax/fillet refs  & Converged after 1 refinement* \\
4  & Box + cylinder union                         & 1           & 22.23  
   & None                      & Success, first attempt \\
5  & Box with cylindrical hole                    & 1           & 23.40  
   & None                      & Success, first attempt* \\
6  & Parametric plate with 4 holes                & 1           & 28.12  
   & None                      & Success, first attempt* \\
7  & Parametric hinge design                      & 3           & 53.53  
   & Geometry constraints     & Converged after 2 refinements* \\
8  & Gear with involute profile                   & 50 (max)    & 836.46 
   & Unsupported API calls    & Did not converge fully \\
9  & Plate with cutouts, chamfers                 & 3           & 81.09  
   & Overconstraints at first & Converged after 2 refinements* \\
10 & Fully constrained rectangular frame          & 50 (max)    & 909.11 
   & Null shape, repeated errs & Did not converge fully \\
\bottomrule
\end{tabular}%
}
\smallskip
\begin{minipage}[b]{\textwidth}
  \footnotesize\textit{Note: * indicates cases where the model generated a design that did not fully align with the user’s provided prompt, though no significant geometric or syntactic errors were found. Deviations from the intended design were observed.}
\end{minipage}
\end{table*}

\noindent

For the simplest tasks (Cases~1 and~2), a single script generation usually sufficed, with execution times under 30~seconds. Slightly more complex designs (Cases~3--5) showed minor geometric or syntax errors that were corrected within one additional refinement loops. However, the generated designs did not fully match the user-provided prompt, reflecting slight deviations from the intended specifications. Cases that demanded parametric constraints or hierarchically dependent features (Cases~6 and~7) required additional adjustments to handle references and geometric consistency, but still converged in under a minute of total runtime. Advanced use cases (Cases~8 and~10) involving specialized geometry (e.g., involute gear profiles) or fully constrained frames triggered repeated failures (up to 50 iterations) and could not successfully converge. Closer inspection of error logs revealed references to unsupported FreeCAD APIs or null-shape issues caused by overconstraint. These persistent errors suggest that current 
LLM-driven prompts lack sufficient built-in context or domain-specific heuristics to navigate highly specialized modeling operations.


Despite these challenges at higher complexities, the overall findings confirm that an iterative, feedback-driven approach can effectively automate CAD scripting for many standard design tasks. By automatically re-prompting with FreeCAD error logs, the system rapidly isolates and corrects most syntax or geometric mistakes. The literature on iterative generative systems similarly emphasizes the importance of incorporating domain-specific knowledge to manage edge cases, constraint logic, and specialized feature libraries. Future refinements in memory-based prompt engineering, combined with deeper integration of FreeCAD’s advanced APIs, may overcome current limitations and further streamline complex parametric workflows. The output of all the experiments are outlined in Appendix ~\ref{sec:exp_results}.

\section{Related Work}
\label{sec:related}
AI in engineering design has been used in multiple works \cite{vardhan2021machine,vardhan2022data}. Most of the earlier work on design generation and optimization relied on Bayesian optimization \cite{vardhan2023constrained} or deep active learning \cite{vardhan2022deepal}.
The integration of Large Language Models (LLMs) into Computer-Aided Design (CAD) is gaining significant attention, particularly for automation, generative design, and error correction \cite{jadhav2024large,wu2023cad}. Research has shown that LLMs can automate CAD tasks by generating parametric models from natural language descriptions \cite{badagabettu2024query2cad} and assisting in repetitive feature generation \cite{zhang2025llm}.  Error detection studies highlight LLMs’ ability to identify and correct syntax and logical errors, reducing debugging time significantly (Patel et al., 2023; Gupta et al., 2022). Open-source CAD tools like FreeCAD \cite{riegel2016freecad} have particularly benefited from LLM-driven automation, with frameworks integrating memory retrieval and iterative refinement for improved script accuracy. While these advancements show promise, challenges remain in maintaining geometric validity, aligning design intent, and optimizing computational efficiency, necessitating further exploration of hybrid AI and geometric reasoning approaches for enhanced CAD automation.


\section{Conclusion}
\label{sec:conclusion}

In this paper, we investigated how LLMs can streamline the generation of FreeCAD scripts for tasks ranging from basic shape creation to advanced parametric modeling. By using an iterative refinement method, simpler geometries and moderately complex assemblies converged successfully, showing that text-based generation can reduce scripting overhead. Failures in specialized or highly constrained geometry, however, highlight important limitations when LLMs must handle advanced or poorly documented procedures.

Despite these hurdles, the results demonstrate that LLM-driven generation, combined with systematic error handling, can substantially minimize manual steps in routine CAD tasks. The iterative approach enables rapid corrections without requiring extensive scripting expertise. These findings confirm the feasibility of using LLMs to automate design workflows and underscore the importance of enhancing model understanding for complex operations.

\section{Future Work}
\label{sec:Futurework}
In future iterations of this system, several refinements can be explored to improve efficiency and robustness. One key enhancement is the implementation of \textbf{LLM memorization techniques} \cite{speicher2024understanding}, where previous FreeCAD scripts, user text input, and their associated error corrections are stored and retrieved using an embedding-based search \cite{peng2023embedding}. This would reduce redundant error correction cycles by allowing the system to recall and apply prior solutions dynamically. Additionally, integrating the pipeline into a cloud-based or containerized environment (e.g., using Docker with a FreeCAD server instance) would enable scalable execution and remote access, making it more adaptable for collaborative design workflows. Further improvements could also involve adaptive prompt engineering \cite{zhou2023llm}, where the LLM dynamically refines the level of specificity in its responses based on recurring failure patterns. Lastly, exploring hybrid AI approaches, such as combining LLM-based script generation with rule-based geometric validation, could enhance the accuracy of the initial script output, reducing reliance on iterative refinements. Given the rapid emergence of new LLMs like DeepSeek, Mistral, and Claude, a holistic comparison across different models to evaluate their efficiency in interacting with FreeCAD could provide valuable insights into selecting the most effective model for CAD automation.

\bibliographystyle{ACM-Reference-Format}
\bibliography{References/references}
\appendices
\onecolumn
\newpage
\appendix

\section{Detailed Prompt Descriptions for FreeCAD Script Generation and Results}
\label{sec:exp_results}

This appendix provides the full set of ten prompts used in our experimental evaluation. These prompts were designed with increasing complexity, from simple geometric shapes to fully constrained parametric designs. 

\begin{table}[h]
    \centering
    \caption{Complexity Scaling of FreeCAD Script Generation}
    \label{tab:prompt_levels}
    \begin{tabular}{|c|l|l|}
        \hline
        \textbf{Level} & \textbf{Model Type} & \textbf{Key Features} \\
        \hline
        1 & Cube & Basic shape, fixed dimensions \\
        2 & Cylinder & Defined radius and height \\
        3 & Filleted Cuboid & Edge fillets, feature modifications \\
        4 & Boolean Union & Merging a box and cylinder \\
        5 & Boolean Subtraction & Cutting a hole through a solid \\
        6 & Parametric Plate & Fully constrained model with drilled holes \\
        7 & Parametric Hinge & Multiple segments, constraints \\
        8 & Gear & Involute profile, precise tooth count \\
        9 & Plate with Cutouts & Complex feature constraints \\
        10 & Parametric Frame & Reinforcement ribs, multiple constraints \\
        \hline
    \end{tabular}
\end{table}

Below is the output image along with the execution time and number of iterations of each task based on the above table designed by FreeCAD with the help of LLM:

\subsection*{Case 1: Basic Cube}
\textbf{Description:} Create a cube in FreeCAD with dimensions length = 50mm, width = 50mm, height = 50mm. Ensure the cube is positioned at the origin (0,0,0) in the global coordinate system.

\textbf{Result:}

Total Execution Time: 19.06 seconds

Number of iterations: 1
\begin{figure}[H] 
    \centering
    \includegraphics[width=0.8\textwidth]{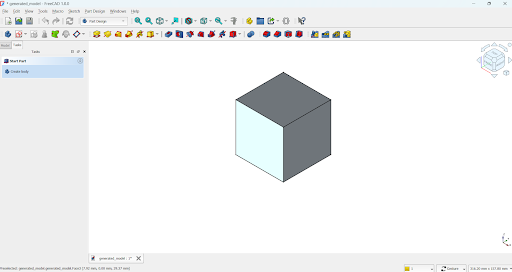}
    \caption{Output for Case 1.}
    \label{fig:output1}
\end{figure}

\subsection*{Case 2: Cylinder}
\textbf{Description:} Create a cylinder with radius = 30mm and height = 80mm. The base of the cylinder should be positioned at (0,0,0) on the XY-plane.

\textbf{Result:}

Total Execution Time: 20.29 seconds

Number of iterations: 1
\begin{figure}[H]
    \centering
    \includegraphics[width=0.8\textwidth]{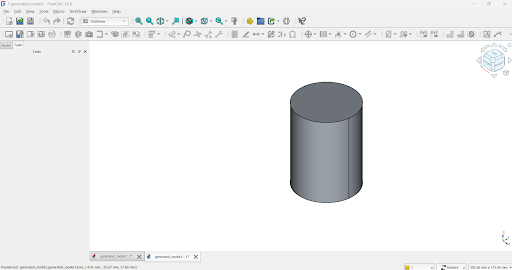}
    \caption{Output for Case 2.}
    \label{fig:output1}
\end{figure}

\subsection*{Case 3: Cuboid with Fillets}
\textbf{Description:} Create a cuboid with dimensions length = 100mm, width = 50mm, height = 30mm. Apply a fillet of 5mm to all edges. Position the cuboid such that its bottom face lies on the XY-plane.

\textbf{Result:}

Total Execution Time: 42.0 seconds

Number of iterations: 2
\begin{figure}[H]
    \centering
    \includegraphics[width=0.8\textwidth]{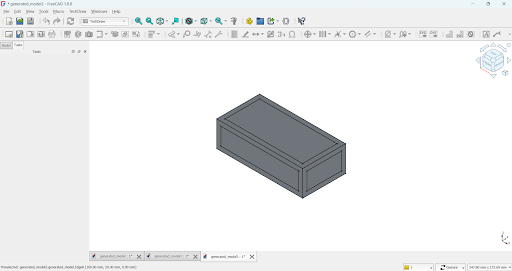}
    \caption{Output for Case 3.}
    \label{fig:output1}
\end{figure}

\subsection*{Case 4: Boolean Union (Box + Cylinder)}
\textbf{Description:} Create a box with dimensions 80mm × 50mm × 30mm and a cylinder with radius 15mm and height 50mm. Position the cylinder such that its base is on the top face of the box, centered along its width. Perform a Boolean union to merge both shapes into a single solid.

\textbf{Result:}

Total Execution Time: 22.23 seconds

Number of iterations: 1
\begin{figure}[H]
    \centering
    \includegraphics[width=0.8\textwidth]{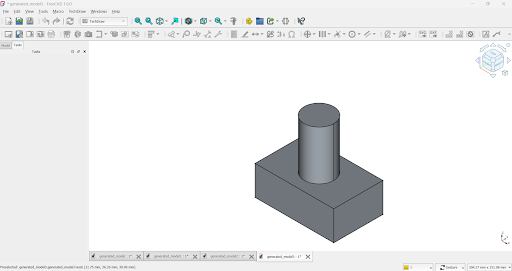}
    \caption{Output for Case 4.}
    \label{fig:output1}
\end{figure}

\subsection*{Case 5: Boolean Subtraction (Cut Cylinder from Box)}
\textbf{Description:} Create a box with dimensions 120mm × 60mm × 40mm and a cylinder with radius 10mm and height 50mm. Position the cylinder such that its base is on the XY-plane and its center aligns with the middle of the box along its width. Perform a Boolean subtraction to cut the cylinder from the box, creating a hole through its height.

\textbf{Result:}

Total Execution Time: 23.40 seconds

Number of iterations: 1
\begin{figure}[H]
    \centering
    \includegraphics[width=0.8\textwidth]{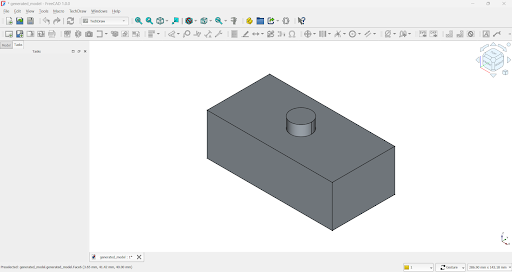}
    \caption{Output for Case 5.}
    \label{fig:output1}
\end{figure}

\subsection*{Case 6: Parametric Plate with Holes}
\textbf{Description:} Create a parametric rectangular plate with dimensions length = 150mm, width = 100mm, thickness = 10mm.
\begin{itemize}
    \item Drill four circular holes of radius 5mm, each hole’s center should be located 10 mm from both adjacent edges of the plate
    \item The holes should be placed near each of the four corners of the rectangle.
    \item Ensure the model is fully constrained so that modifying any dimension updates all features accordingly.
\end{itemize}

\textbf{Result:}

Total Execution Time: 28.12 seconds

Number of iterations: 1
\begin{figure}[H]
    \centering
    \includegraphics[width=0.8\textwidth]{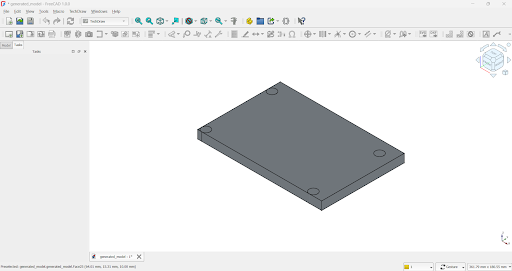}
    \caption{Output for Case 6.}
    \label{fig:output1}
\end{figure}

\subsection*{Case 7: Parametric Hinge Design with Constraints}
\textbf{Description:} Design a parametric hinge with the following specifications:

\begin{itemize}
    \item Leaf 1: Length = 120mm, Width = 30mm, Thickness = 5mm.
    \item Leaf 2: Length = 100mm, Width = 30mm, Thickness = 5mm.
    \item Knuckle: Diameter = 10mm, Length = 30mm (centered between the two leaves).
    \item Pin: Diameter = 8mm, Length = 35mm (extending slightly beyond the knuckle).
\end{itemize}
Constraints: Ensure the hinge can rotate freely around the pin. Apply a chamfer of 2mm to all outer edges of the leaves. Make the model fully parametric, allowing adjustments to the length, width, and thickness of the leaves, as well as the diameter and length of the knuckle and pin, without breaking the design.

\textbf{Result:}

Total Execution Time: 53.53 seconds

Number of iterations: 3
\begin{figure}[H]
    \centering
    \includegraphics[width=0.8\textwidth]{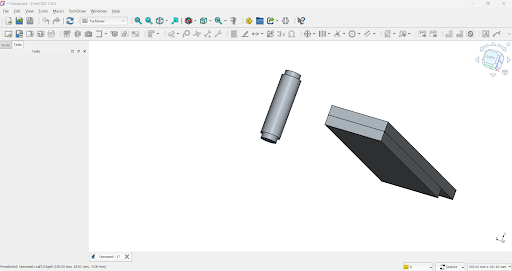}
    \caption{Output for Case 7.}
    \label{fig:output1}
\end{figure}

\subsection*{Case 8: Gear with Custom Tooth Profile}
\textbf{Description:} Create a gear with the following specifications:  
- Outer diameter = 120mm.  
- Inner bore diameter = 20mm.  
- Number of teeth = 24.  
- Tooth profile = involute gear profile.  
Ensure that the gear is centered at the origin and is ready for 3D printing.

\textbf{Result:}

Total Execution Time: 836.46 seconds

Number of iterations: 50

\begin{verbatim}
STDERR:

Exception while processing file: freecad_generated_script.py [module 'Part' 
has no attribute 'makeGear']
\end{verbatim}

\subsection*{Case 9: Plate with Cutouts}
\textbf{Description:} Create a flat metal plate with dimensions 200mm × 150mm × 8mm. Add the following features:  
- Two rectangular cutouts (40mm × 20mm) positioned symmetrically along the plate’s length.  
- Four circular holes (diameter 10mm) positioned at each corner, 15mm from the edges.  
Apply a chamfer of 3mm on all external edges. Ensure the model is parametric so that changes to dimensions update all cutouts and holes proportionally.

\textbf{Result:}

Total Execution Time: 81.09 seconds

Number of iterations: 3
\begin{figure}[H]
    \centering
    \includegraphics[width=0.8\textwidth]{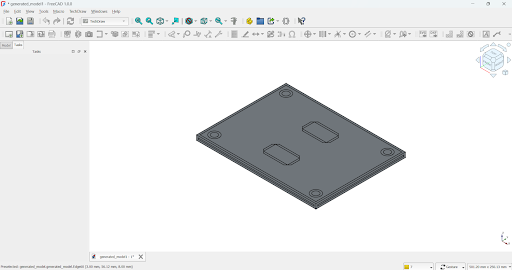}
    \caption{Output for Case 9.}
    \label{fig:output1}
\end{figure}

\subsection*{Case 10: Fully Constrained Parametric Frame}
\textbf{Description:} Create a parametric rectangular frame with an outer dimension of 250mm × 180mm and a wall thickness of 15mm. The frame should have:  
- Rounded corners with a radius of 10mm.  
- Mounting holes (diameter 8mm) at each corner, 20mm from the edges.  
- An internal reinforcement rib, running horizontally and vertically through the center with a thickness of 10mm.  
Ensure that all constraints are defined so that modifying any dimension updates the entire frame proportionally.

\textbf{Result:}

Total Execution Time: 909.11 seconds

Number of iterations: 50
\begin{verbatim}
STDERR:

Exception while processing file: freecad_generated_script.py [Null shape]

\end{verbatim}

\end{document}